\documentstyle[12pt]{article}
\begin{document}

\begin{center}
\Large{ON QUASITRIANGULAR HOPF ALGEBRAS \\ RELATED
TO THE BOREL SUBALGEBRA OF $sl_2$}
\end{center}

\vspace{.5cm}

\begin{center}
\large{A.A.VLADIMIROV}
\end{center}
\begin{center}
{\em Laboratory of Theoretical Physics,
Joint Institute for Nuclear Research, \\
Dubna, Moscow region, 141980, Russia}
\end{center}
\begin{center}
{\em E-mail:} \ alvladim@thsun1.jinr.dubna.su
\end{center}

\vspace{.5cm}

\begin{center}
ABSTRACT
\end{center}

Explicit isomorphism is established between quasitriangular Hopf
algebras studied recently by O.Ogievetsky and by the present author.

\vspace{1.5cm}

In my recent papers~\cite{Vl1,Vl2} a regular method has been proposed
for constructing quasitriangular Hopf algebras (actually, quantum
doubles) out of invertible matrix solutions $R$ of the
quantum Yang-Baxter equation
\begin{equation}
R_{12}R_{13}R_{23}=R_{23}R_{13}R_{12}\,.  \label{1}
\end{equation}
As an illustration, the $R$-matrix of Jordanian type~\cite{Ly,DMMZ}
\begin{equation} \left(
\begin{array}{cccc}1&1&-1&1\\0&1&0&1\\0&0&1&-1\\0&0&0&1 \end{array}
\right)  \label{2}
\end{equation}
has been shown~\cite{Vl2,Vl3} to produce, among others, a
quasitriangular Hopf algebra with generators $\{v,h\}$ obeying the
relations
$$ [v,h]=2\sinh h\,, $$
\begin{equation}
\Delta(v)=e^h\otimes v+v\otimes e^{-h}\,,\ \ \
\Delta(h)=h\otimes 1+1\otimes h\,,  \label{3}
\end{equation}
$$ S^{\pm1}(h)=-h\,,\ \ \ S^{\pm1}(v)=-v\mp2\sinh h\,. $$
Its universal ${\cal R}$-matrix is found~\cite{Vl3} to be
\begin{equation} {\cal
R}=\exp\left\{\frac{h\otimes 1+1\otimes h} {\sinh(h\otimes 1+1\otimes
h)}(\sinh h\otimes v-v\otimes \sinh h) \right\}\,. \label{4}
\end{equation}

After submitting the paper~\cite{Vl3} I have learned from O.Ogievetsky
that in paper~\cite{Og}, using quite different approach (which allows
classification of all Hopf structures on the Lie algebra $[x,h]=2h$),
he obtained a quasitriangular Hopf algebra $\{\tau ,\sigma \}$ defined
by $$ [\tau ,\sigma ]=2(1-e^\sigma )\,, $$
\begin{equation}
\Delta(\tau )=\tau \otimes e^\sigma +1\otimes \tau \,, \ \ \ \
\Delta(\sigma )=\sigma \otimes 1+1\otimes \sigma \,, \label{5}
\end{equation}
$$ S(\tau )=-\tau e^{-\sigma }, \ \ S^{-1}(\tau )=-\tau e^{-\sigma }
+2(1-e^{-\sigma }), \ \ S^{\pm 1}(\sigma )=-\sigma , $$
with universal ${\cal R}$-matrix
\begin{equation}
{\cal R}=\exp(\frac{\sigma }{2}\otimes \tau )\,\exp(-\tau \otimes
\frac{\sigma }{2})\,.  \label{6}
\end{equation}
Since the Hopf algebra (\ref{3}), due to a reparametrization
\begin{equation}
v=\frac{\sinh h}{h}x \ \ \ \ \Longrightarrow \ \ \ \ [x,h]=2h\,,
\label{7}  \end{equation}
evidently belongs to the class considered by Ogievetsky, it is only
natural to look for an isomorphism between the Hopf algebras (\ref{3})
and (\ref{5}).

Such an isomorphism really exists and can be fixed by
\begin{equation}
\tau =-e^{-h}v\,, \ \ \ \ \ \sigma =-2h\,.  \label{8}
\end{equation}
Relations (\ref{5}) are readily obtained from (\ref{3}) and (\ref{8}).
As a by-product, we come to an alternative form of the $\cal R$-matrix
(\ref{4}):
\begin{equation}
{\cal R}=\exp(h\otimes e^{-h}v)\,\exp(-e^{-h}v\otimes h)\,.\label{4.1}
\end{equation}

In~\cite{Vl2,Vl3} (see also~\cite{BH}), the following quasitriangular
Hopf algebra $\{b,g,v,h\}$ (the quantum double of $\{v,h\}$) was also
considered:
$$ [g,b]=[h,b]=2\sinh g\,,\ \ \
[g,v]=[h,v]=-2\sinh h\,, $$ $$ [b,v]=2(\cosh g)v+2(\cosh h)b\,,\ \ \ \
[g,h]=0\,, $$ \begin{equation} \Delta(b)=e^g\otimes b+b\otimes
e^{-g}\,,\ \ \ \Delta(v)=e^h\otimes v+v\otimes e^{-h}\,, \label{9}
\end{equation}
$$ \Delta(g)=g\otimes 1+1\otimes g\,,\
\Delta(h)=h\otimes 1+1\otimes h\,,\ S^{\pm1}(g)=-g\,, $$
$$ S^{\pm1}(h)=-h\,,\ S^{\pm1}(b)=-b\pm2\sinh g\,,\
S^{\pm1}(v)=-v\mp2\sinh h\,. $$
The (anti)duality relations between $\{b,g\}$ and $\{v,h\}$ are:
$$ <1,b>=<1,g>=<v,1>=<h,1>=0\,, $$
\begin{equation}
<1,1>=<h,b>=<v,g>=1\,, \label{10}
\end{equation}
$$ <v,b>=-1\,, \ \ \ \ <h,g>=0\,. $$
Universal $\cal R$-matrix looks like~\cite{Vl3}
\begin{equation} {\cal
R}=\exp\left\{\frac{g\otimes 1+1\otimes h} {\sinh(g\otimes 1+1\otimes
h)}(\sinh g\otimes v+b\otimes \sinh h) \right\}\,. \label{11}
\end{equation}
In terms of $\{\tau ,\sigma \}$ and their antiduals $\{\mu ,\nu \}$ the
same quantum double is characterized by the relations
$$ [\tau ,\sigma ]=[\tau ,\nu ]=2(1-e^\sigma )\,, \ \
[\mu ,\sigma ]=[\mu ,\nu ]=2(1-e^\nu )\,, $$
$$ [\tau ,\mu ]=2(\mu -\tau )\,, \ \ \ \ \ [\sigma ,\nu ]=0\,, $$
$$ \Delta(\tau )=\tau \otimes e^\sigma +1\otimes \tau \,, \ \
\Delta(\sigma )=\sigma \otimes 1+1\otimes \sigma \,, $$
\begin{equation}
\Delta(\mu )=\mu \otimes e^\nu +1\otimes \mu \,, \ \
\Delta(\nu )=\nu \otimes 1+1\otimes \nu \,,  \label{12}
\end{equation}
$$ S(\tau )=-\tau e^{-\sigma }\,, \ \ S^{-1}(\tau )=-\tau e^{-\sigma }
+2(1-e^{-\sigma })\,, $$
$$ S(\mu )=-\mu e^{-\nu }\,, \ \ S^{-1}(\mu )=-\mu e^{-\nu }
+2(1-e^{-\nu })\,, $$
$$ S^{\pm 1}(\sigma )=-\sigma \,, \ \ \ \ \ S^{\pm 1}(\nu )=-\nu  $$
with antiduality conditions
$$ <1,\mu >=<1,\nu >=<\tau ,1>=<\sigma ,1>=0\,, $$
\begin{equation}
<1,1>=1\,, \ \ \ <\tau ,\nu >=2\,, \ \ \ <\sigma ,\mu >=-2\,,\label{13}
\end{equation}
$$ <\tau ,\mu >=<\sigma ,\nu >=0 $$
and universal $\cal R$-matrix~\cite{Og}
\begin{equation}
{\cal R}=\exp(\frac{\nu }{2}\otimes \tau )\,\exp(-\mu \otimes
\frac{\sigma }{2})\,. \label{14}
\end{equation}
Apparent similarity of (\ref{14}) and (\ref{6}), as well as (\ref{11})
and (\ref{4}), is due to 'selfduality'~\cite{Vl3,Og} of both
$\{\tau ,\sigma \}$- and $\{v,h\}$-algebras.

Hopf algebras (\ref{9}) and (\ref{12}) are isomorphic by
\begin{equation}
\mu =e^{-g}b\,, \ \ \nu =-2g\,, \ \ \tau =-e^{-h}v\,, \ \
\sigma =-2h\,. \label{15}
\end{equation}
In particular, this produces an alternative representation for $\cal R$
(\ref{11}):
\begin{equation}
{\cal R}=\exp(g\otimes e^{-h}v)\,\exp(e^{-g}b\otimes h)\,. \label{11.1}
\end{equation}
Surprisingly, it appears problematic to derive (\ref{11}) directly from
(\ref{11.1}) by the mere use of Baker-Campbell-Hausdorff formula.

\vspace{.5cm}

\noindent {\bf Acknowledgments}

\vspace{.3cm}

The present work was supported by the Heisenberg-Landau program, 1993.

\vspace{.3cm}

I am very grateful to Prof. J.Wess for organizing my visit to Munich
where this research was carried out. I appreciate stimulating
discussions with O.Ogievetsky, A.Kempf, M.Pillin and R.Engeldinger.

\end{document}